\documentclass[epj]{svjour}
\usepackage{graphicx}
\usepackage{amsmath}
\usepackage{amssymb}
\usepackage{epsfig}

\newcommand{\cE} {{\cal E}}
\newcommand{\cG} {{\cal G}}
\newcommand{\Deltab} {\Delta_{\rm b}}
\newcommand{\Deltac} {\Delta_{\rm c}}

\newcommand{\half} {\frac{1}{2}}
\newcommand{\ie} {{\it i.e.}, }
\newcommand{\kc} {k_{\rm c}}
\newcommand{\Order}{{\cal O}}
\newcommand{\Pc} {P_{\rm c}}
\newcommand{\pd} {\partial}
\newcommand{\rmd} {{\rm d}}

\begin{document}

\thispagestyle{empty}

\centerline{\bf Comment}

This manuscript was not published eventually, for two reasons:
\begin{itemize}
  
\item[$\bullet$] The findings were too similar to the localization of
  wrinkles obtained earlier for a sheet on an elastic foundation. See
  G.\ W.\ Hunt, M.\ K.\ Wadee, and N.\ Shiacolas, J.\ Appl.\ Mech.\
  {\bf 60}, 1033 (1993).  Note, however, that the two problems are not
  equivalent. For an elastic foundation the second term in the energy
  density, eq.\ (1), is $(K/2)h^2$ rather than $(K/2)h^2\cos\phi$.
  
\item[$\bullet$] We have found the exact solution to the problem
  presented here.  See arXiv:1107.5505.

\end{itemize}

In addition, B.\ Audoly has followed this preprint to show, using an
amplitude-equation approach, that the variational {\it Ansatz} used
here is indeed the profile minimizing the energy to quartic terms in
the height amplitude. See B.\ Audoly, Phys.\ Rev. E {\bf 84}, 011605
(2011). This article also presents a unified perturbative treatment
for several types of substrate, including the fluid substrate and the
linear elastic foundation.

If you find the manuscript useful nonetheless, and wish to cite it,
please use arXiv:1009.2487.

%============================================================
\title{Instability of infinitesimal wrinkles against folding}
%============================================================

\author{Haim Diamant\inst{1} \and Thomas A.\ Witten\inst{2}} 

\institute{Raymond and Beverly Sackler School of Chemistry, 
Tel Aviv University, Tel Aviv 69978, Israel; 
\email{hdiamant@tau.ac.il} 
\and
Department of Physics and James Franck Institute,
University of Chicago, Chicago, Illinois 60637, USA;
\email{t-witten@uchicago.edu}
}

\date{\today}

\abstract{ 
  We analyze the buckling of a rigid thin membrane floating on a
  dense fluid substrate.  The interplay of curvature and substrate
  energy is known to create wrinkling at a characteristic wavelength
  $\lambda$, which localizes into a fold at sufficient buckling
  displacement $\Delta$.  By analyzing the regime $\Delta\ll\lambda$,
  we show that wrinkles are unstable to localized folding for {\em
    arbitrarily small} $\Delta$.  After observing that evanescent
  waves at the boundaries can be energetically favored over uniform
  wrinkles, we construct a localized {\it Ansatz} state far from
  boundaries that is also energetically favored.  The resulting
  surface pressure $P$ in conventional units is $2-(\pi^2/4)
  (\Delta/\lambda)^2$, in entire agreement with previous numerical
  results. The decay length of the amplitude is
  $\kappa^{-1}=(2/\pi^2)\lambda^2/\Delta$.  This case illustrates how
  a leading-order energy expression suggested by the infinitesimal
  displacement can give a qualitatively wrong configuration.
}

\maketitle

%------------------------------------------------

%=====================
\section{Introduction}
%=====================
\label{sec_intro}

Recent work has revealed a wealth of spontaneous spatial structures in
thin elastic sheets, owing to their easy deformability by bending
\cite{Bowden1998,Sharon:2002kx,CerdaMaha,Cerda:2004dp,Gopal:2006eu,Witten:2007ai,Huang:2007vn,Audoly:2008ys,Lee:2008qe,LukaCerda,Py:2009uq,BennyPRE,Huang:2010sp,Holmes:2010fk,Leahy:2010zr}.
Special interest has focusses on the recently-proposed wrinkle-to-fold
transition in compressed thin sheets floating on a dense liquid
\cite{LukaCerda}.  Such composite structures occur widely in
industrial coatings and biological tissues.  The sheet can be a
50-nm-thick metal film \cite{Audoly:2008ys} or a two-nanometer-thick
lipid monolayer \cite{Lee:2008qe}.

When compressed to the point of buckling, the incipient deformation is
an undulation or wrinkle whose wavelength $\lambda$ is a combination
of the bending stiffness and the substrate rigidity, as specified
below.  A compressional displacement $\Delta$ accompanies this
buckling.  Buckling occurs when the compressional force per unit
length $P$ exceeds a threshold.  In the conventional units defined
below, this threshold $P$ is equal to 2.  As with the classic Euler
buckling of a compressed rod \cite{landau:44}, the buckled structure is
unstable under this force: larger displacement creates more bending
deflection, which requires a smaller pressure.  Thus $P$ decreases as
$\Delta$ increases.

For sufficiently large $\Delta$ of order $\lambda$, the
weak-deflection expression for the energy becomes inadequate, and the
selected structure is no longer a uniform undulation but a localized
fold.  Reference \cite{LukaCerda} explored the transition from the
wrinkled to the folded state numerically and experimentally,
concluding that the transition occurs for $\Delta\simeq\lambda/3$.
They determined numerically that the dimensionless force $P$ is
well represented as $2-2.47(\Delta/\lambda)^2$.  As $\Delta$
decreases, the buckled region was observed to be less and less
localized, and to span more and more wavelengths, eventually spanning
the system.

These observations led us to consider the regime of indefinitely small
displacements $\Delta\ll\lambda$, in order to discern how the
corrections to the small-deformation energy emerge.  As explained
below, we found that these corrections remain important even for
arbitrarily small displacements.  The resulting buckling pattern is
always localized, though its spatial extent diverges as the
displacement $\Delta$ decreases to zero.  Since the regime of small
displacements is analytically tractable, we could obtain a closed-form
expression for the $P(\Delta)$ function: $P = 2 - (\pi^2/4)
(\Delta/\lambda)^2 + \Order((\Delta/\lambda)^4)$. This agrees
completely with the numerical value cited above. We also obtain a
closed-form expression for the buckling profile in this regime.
Though this form is obtained by a variational {\it Ansatz}, we argue
below that it approaches the exact profile in the limit of small
displacement.

%==============
\section{Model}
%==============
\label{sec_model}

\begin{figure}[tbh]
\vspace{0.5cm}
\centerline{\resizebox{0.48\textwidth}{!}
{\includegraphics{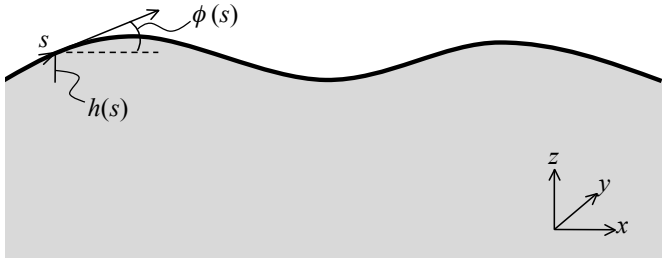}}}
\caption[]{Schematic view of the system and its parametrization.}
\label{fig_scheme}
\end{figure}

We consider a two-dimensional problem of a thin incompressible elastic
sheet of length $L$, width $W$, and bending modulus $B$. The sheet is assumed to deform in the
$xz$ plane while remaining uniform along the $y$ axis. Being
incompressible, its total length $L$ is fixed, and its configuration,
including the total displacement along the $x$ axis, $\Delta$, is
fully accounted for by a height profile, $h(s)$, as a function of
arclength $s\in[-L/2,L/2]$.  The region $z<h$ is occupied by a
liquid of mass density $\rho=K/g$ ($g$ being the gravitational
acceleration). See fig.\ \ref{fig_scheme}.

For a certain configuration, $h(s)$, the energy of the sheet $E$ and
the displacement $\Delta$ are given by
\begin{eqnarray}
  E &=& W\int_{-L/2}^{L/2} \rmd s
  \left( \half B\dot{\phi}^2 + \half K h^2\cos\phi \right),
 \nonumber\\
\label{E1}
  \Delta &=& \int_{-L/2}^{L/2} \rmd s 
  \left( 1 - \cos\phi \right), 
%\label{Delta1} 
\end{eqnarray} 
where $\phi(s)$ is the angle between the local tangent to the sheet
and the $x$ axis at arclength $s$ (cf.\ fig.\ \ref{fig_scheme}), and a
dot denotes a derivative with respect to $s$.  It is helpful to work
in units where $B=K=1$, \ie rescale the energy by $B$ and all lengths
by $(B/K)^{1/4}$. The pressure $P$ is then scaled by $(BK)^{1/2}$.  We
shall also let $W=1$ in what follows.  Using these definitions and the
geometrical relation $\dot{h}=\sin\phi$, we rewrite eq.\ (\ref{E1}) %and (\ref{Delta1}) 
as
\begin{eqnarray} 
  E &=& \half \int_{-L/2}^{L/2} \rmd s \left(
  \frac{\ddot{h}^2}{1-\dot{h}^2} +
  h^2\sqrt{1-\dot{h}^2} \right),
 \nonumber\\
\label{E2}
  \Delta &=& \int_{-L/2}^{L/2} \rmd s
  \left( 1-\sqrt{1-\dot{h}^2} \right).
%\label{Delta2}
\end{eqnarray}

Our aim is to find the height profile, $h(s)$, which minimizes the
energy of the sheet for a given displacement $\Delta$, 
$E=E^*(\Delta)$; from this we may find the resulting
pressure--displacement relation,  $P(\Delta)=\rmd E^*/\rmd\Delta$.

\section{Extended wrinkling}
%---------------------------
\label{sec_wrinkle}

Let us substitute in eq.\ (\ref{E2}) a uniformly wrinkled height
profile of amplitude $A$ and wavenumber $k$,
\begin{equation}
  h(s) = A\cos(ks).
\label{wrinkle}
\end{equation}
The extensivity of the energy implies that $E=L\cE(\delta)$ with
$\delta=\Delta/L$. To second order in $A^2$ we get
\begin{eqnarray}
  \cE &=& \frac{1}{4}(k^4+1)A^2 + \frac{1}{32}k^2(2k^4-1)A^4,
\label{cE1} \\
  \delta &=& \frac{1}{4}k^2A^2 + \frac{3}{64}k^4A^4.
\label{delta1}
\end{eqnarray}

We first examine the leading order. Substituting eq.\ (\ref{delta1})
in eq.\ (\ref{cE1}) yields $\cE(\delta)=(k^2+k^{-2})\delta$. For any
$\delta>0$ the energy has a minimum of $\cE^*=2\delta$ for $k=1$. This
marks the linear wrinkling instability
\cite{Milner,CerdaMaha,ZhangWitten}, with a finite wavenumber $\kc=1$
[wavelength $\lambda=2\pi(B/K)^{1/4}$], squared amplitude
$A^2=4\delta$, and energy $E^*=2\Delta$. The resulting critical
pressure is
$
  \Pc = \rmd E^*/\rmd\Delta = 2
$.

Repeating the same procedure in the next order, we find from eqs.\ 
(\ref{cE1}) and (\ref{delta1})
\begin{equation}
  \cE(\delta) = (k^2+k^{-2})\delta + (k^2-5k^{-2})\delta^2/4.
\end{equation}
This expression is still minimized by $k=\kc=1$, yielding $\cE^*(\delta)=
2\delta-\delta^2$, \ie
\begin{eqnarray}
  E^*(\Delta) &=& 2\Delta - \Delta^2/L \nonumber\\
  P(\Delta) &=& 2 - 2\Delta/L.
\label{E*wrinkle}
\end{eqnarray}

\section{Localized folding}
%--------------------------
\label{sec_fold}

While the solution to the leading-order problem must be of the
wrinkling form [eq.\ (\ref{wrinkle})], such an extended profile does
not necessarily minimize the anharmonic energy.  Indeed, the
next-leading energy correction in eq.\ (\ref{E*wrinkle}) vanishes as
$L\rightarrow\infty$; the wrinkle profile gains no energetic benefit
from this correction in the infinite system.

To see that a localized solution should be favorable, let us fix the
pressure (rather than the displacement) at a value slightly below
threshold, $P=2-p$ with $0<p\ll 1$, and re-examine the leading-order
problem. The energy to be minimized in this case is
$\cG=\cE-P\delta\simeq A^2(k^4-Pk^2+1)/4$, where we have used the
leading order in eqs.\ (\ref{cE1}) and (\ref{delta1}). The most
favorable wrinkling profile has $k^2=1$ and $\cG=(A^2/4)p>0$, which is
larger than $\cG=0$ of the flat state. Thus, the flat state is
linearly stable against wrinkling for $P<\Pc$. Nonetheless, we can get
a lower energy if we allow for a complex wavenumber. We may obtain
$\cG=0$ by taking
\begin{equation}
  k \simeq 1 \pm i\kappa,\ \ \kappa = \sqrt{p}/2.
\label{evanescent}
\end{equation}
Hence, according to the leading-order theory, if there were a boundary at
$s=0$, an `evanescent wave' adjacent to the boundary, of the form
$h(s>0)=Ae^{is - \kappa s}$, could be stabilized for $P<\Pc$. For such
a fixed pressure the flat state is linearly stable against wrinkling,
and a `propagating wave' can be stabilized strictly at $P=\Pc$.

We are interested, however, in profiles which are localized far away
from boundaries. Accordingly, we seek a wave-packet profile with a
spectral content similar to the evanescent-wave profile above. One
such profile is
\begin{equation}
  h(s) = A \frac{\cos(ks)}{\cosh(\kappa s)},
\label{ansatz}
\end{equation}
where the decay coefficient $\kappa$ is used as a variational
parameter, and we take $k=\kc=1$. (Letting $k$ be another variational
parameter does not change the results.) We assume the limit
$L^{-1}\ll\kappa\ll k=1$, where the wavy profile is `underdamped' but
decays sufficiently fast for the system size to be taken as infinite.
The extensivity of the energy again implies
$E=\kappa^{-1}\cE(\delta)$, where now $\delta=\kappa\Delta$. As in the
preceding section we perform the calculation to order $\delta^2$; yet,
unlike the extended-wrinkling case, the result will be of higher order
in $\Delta$, since the domain size itself, $\kappa^{-1}$, depends on
$\Delta$.

Substituting eq.\ (\ref{ansatz}) in eq.\ (\ref{E2}), %and (\ref{Delta1}) 
we obtain within the assumed order of approximation,
\begin{eqnarray}
  \cE &=& (1+\kappa^2)A^2 + \frac{1}{24} A^4
\label{cE2}  \\
  \delta &=& \frac{1}{6}(3+\kappa^2)A^2 + \frac{1}{16}A^4.
\label{delta2}
\end{eqnarray}
Solving eq.\ (\ref{delta2}) for $A^2$,
\begin{equation}
  A^2 = 2(1-\kappa^2/3)\delta - \delta^2/2,
\label{A2fold}
\end{equation}
and substituting it back in eq.\ (\ref{cE2}), we get
$
  \cE(\delta) = 2(1+2\kappa^2/3)\delta - \delta^2/3 
$.
To properly minimize the energy with respect to $\kappa$ for fixed
$\Delta$, we rewrite this result as
\[
  E = 2(1+2\kappa^2/3)\Delta - \kappa\Delta^2/3.
\]

As expected, to the leading order the energy is minimum for $\kappa=0$
and coincides with that of the wrinkling case, $E^*=2\Delta$.
However, with the anharmonic term the energy is minimized by
$\kappa=\Delta/8$, yielding
\begin{eqnarray}
  E^*(\Delta) &=& 2\Delta - \frac{\Delta^3}{48} 
 \nonumber\\
  P(\Delta) &=& 2 - \frac{\Delta^2}{16}.
\label{E*fold}
\end{eqnarray}
Note that the result $\kappa=\Delta/8=\sqrt{2-P}/2$ is identical to
the one obtained for the evanescent wave from the leading-order
problem, eq.\ (\ref{evanescent}).

Finally, substituting $\kappa=\Delta/8$ back in eqs.\ (\ref{A2fold})
and (\ref{ansatz}), we obtain for the localized profile, 
\begin{equation}
  h(s) = \frac{\Delta}{2} \left(1-\frac{7\Delta^2}{384}\right)
  \frac{\cos s}{\cosh[(\Delta/8)s]}.
\label{hfold}
\end{equation}

Equations (\ref{E*fold}) and (\ref{hfold}) are our central results.
Comparing the energies of the extended and localized profiles, eqs.\ 
(\ref{E*wrinkle}) and (\ref{E*fold}), we find that the localized fold
is stabilized for $\Delta>\Deltac=48/L$. Thus, the wrinkle-to-fold
transition presented in ref.\ \cite{LukaCerda} is a finite-size
effect; in the limit $L\rightarrow\infty$  the deformation of the sheet
is always localized in a folded domain of negligible width compared to
the system size, $\kappa^{-1}=8/\Delta$ [or, in dimensional terms,
$\kappa^{-1}=(2/\pi^2)\lambda^2/\Delta$].  The parabolic
pressure--displacement relation of eq.\ (\ref{E*fold}) is rewritten in
dimensional terms as
\[
  \frac{P}{(BK)^{1/2}} = 2 - \frac{\pi^2}{4}
  \left(\frac{\Delta}{\lambda}\right)^2,
\]
which agrees nicely with the numerical result of ref.\ 
\cite{LukaCerda}, $P/(BK)^{1/2}\simeq 2-2.47(\Delta/\lambda)^2$.

From eqs.\ (\ref{E*wrinkle}) and (\ref{E*fold}) one might conclude
that the appearance of the localized state at $\Delta=\Deltac$ is a
first-order transition with discontinuous jumps in $\kappa$ and
$P$. This is an artifact, arising from the assumption $\kappa^{-1}\ll
L$, which breaks down for $\Delta\sim\kappa\sim L^{-1}$. As shown in
the Appendix, a more careful analysis for such small displacements
yields a second-order transition, with $\kappa=0$ for $\Delta<\Deltac$
and $\kappa\sim(\Delta-\Deltac)^{1/2}$ for $\Delta > \Deltac$, leading
to a discontinuous jump in $\rmd P/\rmd\Delta$.

\section{Numerical solution}
%---------------------------

The calculations given in sects.\ \ref{sec_wrinkle} and \ref{sec_fold}
are valid sufficiently close to the instability. To examine the
deformation of the sheet at larger displacements we derive the
Euler-Lagrange equation, corresponding to the minimization of $E$ for a
given $\Delta$, and solve it numerically.

We would like to find the profile $h(s)$ that minimizes the elastic
energy $E$ under the constraint of fixed displacement $\Delta$, where
$E[h(s)]$ and $\Delta[h(s)]$ have been defined in eq.\ (\ref{E2}).
The following equation is obtained from calculating the variation of
$G=E-P\Delta$ with respect to $h(s)$ and setting it to zero:
\begin{eqnarray}
  &&(1-\dot{h}^2)^2 \ddddot{h} + 4(1-\dot{h}^2) \dot{h} \ddot{h} \dddot{h}
  + (1+3\dot{h}^2)\ddot{h}^3 \nonumber\\
  &&+ (h^2/2+P)(1-\dot{h}^2)^{3/2}\ddot{h} 
  + h(1-\dot{h}^2)^{5/2} = 0.
\label{EL}
\end{eqnarray}
This equation needs to be supplemented by four boundary conditions,
for example, the hinge conditions $h(\pm L/2)=0$ and $\ddot{h}(\pm
L/2)=0$.

\begin{figure*}[tbh]
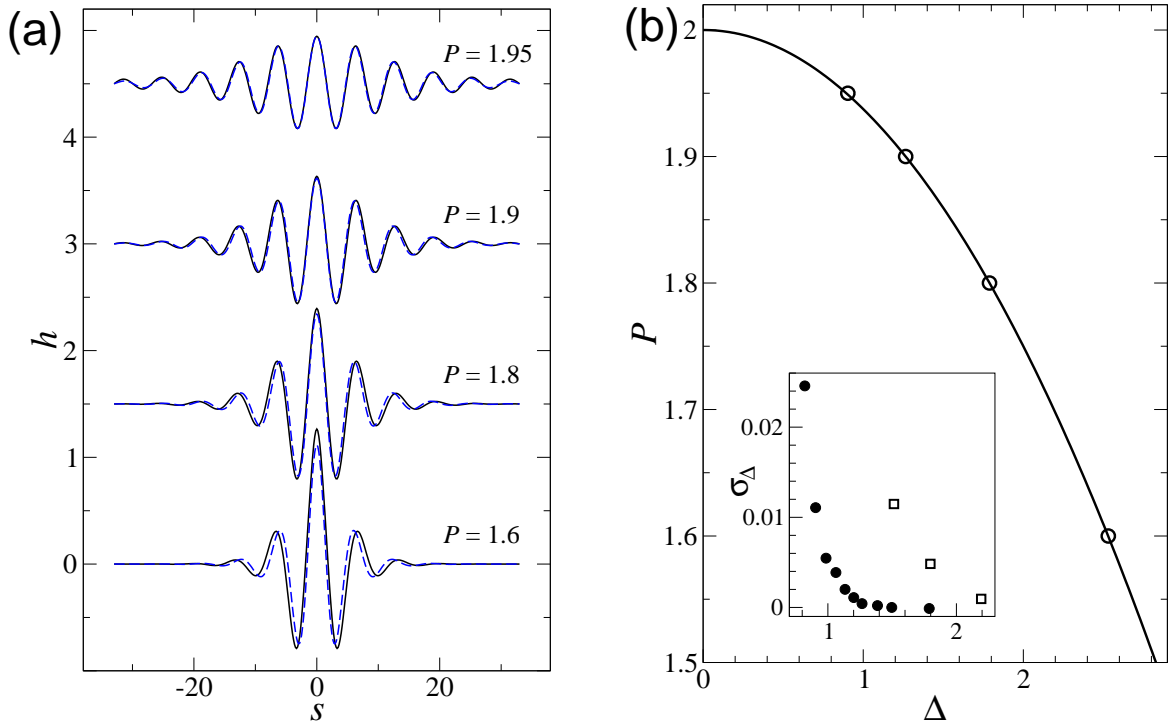

\vspace{0.7cm}
\centerline{\resizebox{0.4\textwidth}{!}
{\includegraphics{fig2a}}
\hspace{0.7cm}
\resizebox{0.4\textwidth}{!}
{\includegraphics{fig2b}}}
\caption[]{(a) Height profiles obtained from the numerical solution
  (black solid curves) and the approximate theory [eq.\ (\ref{hfold}),
  blue dashed curves]. Different curves correspond (from top to
  bottom) to decreasing pressure (as indicated) and increasing
  displacement. The length is $L=21\pi$. Consecutive curves are
  vertically shifted by $1.5$ for clarity. Note that the profiles are
  presented in terms of the material coordinate $s$ rather than the
  projected coordinate $x$.  (b) Pressure as a function of
  displacement. The open circles are the numerical values obtained
  from the profiles shown in (a). The solid curve is the theoretical
  prediction [eq.\ (\ref{E*fold})]. The inset shows the normalized
  discrepancy $\sigma_\Delta$ between our numerical $\Delta(P)$ and
  that of eq.\ (\ref{E*fold}); specifically, $\sigma_\Delta\equiv
  (\Delta_{\rm num}- \Delta_{\rm th})/\Delta_{\rm num}$. Solid circles
  are for $L=21\pi$; open squares are for $L=11\pi$.}
\label{fig_numeric}
\end{figure*}

In the following examples we look for even profiles, $h(-s)=h(s)$,
over a system length of $L=21\pi$ (\ie containing $10\half$
wrinkling wavelengths). For a given $P<\Pc=2$, eq.\ (\ref{EL}) is
solved numerically over the interval $s\in(-L/2,0)$ using the
following boundary conditions: $h(-L/2)=0$, $\ddot{h}(-L/2)=0$,
$\dot{h}(0)=0$, $h(0)=h_0$. We shoot for the value of $h_0$ so as to
make $|\dddot{h}(0)|$ vanish within the required accuracy. Once $h(s)$
is found, the displacement $\Delta(P)$ is calculated using eq.\
(\ref{E2}). The results are presented in fig. \ref{fig_numeric}, where
they are compared with the analytical ones, eqs.\ (\ref{E*fold}) and
(\ref{hfold}).

\begin{figure}[tbh]
\vspace{0.3cm}
\centerline{\resizebox{0.45\textwidth}{!}
{\includegraphics{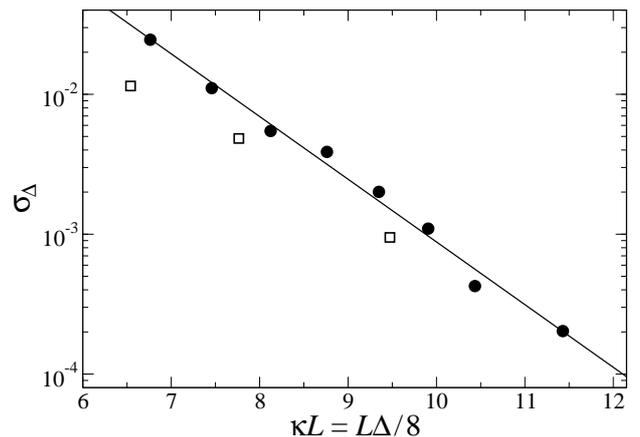}}}
\caption[]{Discrepancy between theory and numerics as a function
 of $\kappa L=(\Delta/8)L$ for $L=21\pi$ (solid circles) and
 $L=11\pi$ (open squares) on a semi-logarithmic scale. The fitted
 line has a slope of $-1.03\pm 0.04$.}
\label{fig_discrepancy}
\end{figure}

The discrepancy between the numerically determined height profile and
the theoretical one is of the order of a few percent, up to about
10\%. It increases with $\Delta$ as expected [fig.\ 
\ref{fig_numeric}(a)]. The predicted parabolic law for $P(\Delta)$,
however, agrees with the numerics to a much greater extent [fig.\ 
\ref{fig_numeric}(b)]. The agreement is remarkable for large
displacements, where the perturbative theory is not expected to hold
at all.  Since we have carried out the analytical calculation to order
$\delta^2$, we should have expected a correction to $E$ of order
$\kappa^{-1}\delta^3 \sim \Delta^5$, leading to a correction of order
$\Delta^4$ in $P$.  Even for nearly vertical deflections (right side
of fig. \ref{fig_discrepancy}) the discrepancy is consistent with
numerical error.  The large discrepancies at small $\Delta$ come from
the finite-$L$ effect discussed above.  This effect should become
exponentially small in $\kappa L\sim L\Delta$, which explains the
sharp decay of the discrepancy $\sigma_\Delta$ with $\Delta$ and its
strong dependence on $L$. [See fig.\ \ref{fig_numeric}(b) inset.]  We
have replotted in fig.\ \ref{fig_discrepancy} the data of that inset
as a function of $\kappa L=(\Delta/8)L$ on a semi-logarithmic scale,
confirming the exponential decay of the discrepancy with $\kappa L$.
We note that the numerical results presented in ref.\ \cite{LukaCerda}
seem to indicate that the parabolic pressure--displacement relation
remains accurate up to the point of self-contact. All of the above
suggests that $P(\Delta)$ of eq.\ (\ref{E*fold}) might, in fact, be
the {\em exact} relation in the limit $\kappa L\gg 1$.

\section{Discussion}
%===================
\label{sec_discuss}

In this section we argue that for the limit of small amplitude our
{\it Ansatz} should become exact. We discuss the mean\-ing of the simple
large-amplitude be\-havior noted above.  We comment on the relevance of
our result for recent Langmuir buckling observations.  We draw
parallels to other continuous instabilities that are not predicted by
lowest-order perturbation of the energy function.

Figure \ref{fig_numeric} shows a near match between the numerically
determined profile and the {\it Ansatz} of eq.\ 
(\ref{ansatz}).  The agreement becomes better as the threshold is
approached.  In this limit our asymptotic $\Order(A^4)$ energy becomes
accurate.  Further, the spectral profile of eq.\ (\ref{ansatz})
approaches that of the evanescent wave that minimizes this asymptotic
energy.  This profile has a peak at $k = 1$ of width $\kappa$.  The
profile of eq.\ (\ref{ansatz}) has a spectral weight that falls off
exponentially for $k-1\gg\kappa$.  We expect that any {\it Ansatz}
with such a spectral distribution would approach the actual profile
near the threshold.  However, not any profile with the proper width and peak
position is adequate.  One such profile is $h(s)=A\sin s ~e^{-\kappa
  |s|}$.  This profile gives the same qualitative
behavior of the energy and pressure for a given displacement $\Delta$
as eq.\ (\ref{ansatz}), yet, here the spectral weight falls off as a
power of $k-1$, rather than the exponential falloff of eq.\ 
(\ref{ansatz}).  Accordingly, we found that its energy is fractionally
larger than that of eq.\ (\ref{ansatz}).  

Our {\it Ansatz} is not unique.  For example, the alternative form 
\[
h(s) = A \cos(ks) / \cosh^2(\kappa s/2)
\]
has the same spectral features noted for eq. (\ref{ansatz}).  However,
neither profile is exact; thus when expanded in powers of $\Delta$,
both must differ from the true behavior in some order.  For example,
the $(7/384)\Delta^2$ correction in eq.\ (\ref{hfold}) depends on our
choice of {\it Ansatz}.

The theory predicts a quadratic initial dependence of pressure on
displacement, as reported already in ref.\ \cite{LukaCerda}.  As noted
above, the quadratic dependence remains accurate far beyond the
small-displacement regime where it was derived.  This accuracy
suggests that the system may be described in a way that is manifestly
homogeneous in $\Delta$. However, we have not been able to find a
simple theory of this type.

The main experimental implication of the theory above is that one
should not expect to see a transition from a wrinkled to a folded
state in a large experimental system. For overall displacements
$\Delta$ of order $\lambda$ this finding is implicit in the work of
ref. \cite{LukaCerda}.  The strain and the wrinkle amplitude at the
threshold of their fold behavior goes to zero as the system size $L$
goes to infinity. Nevertheless, our work shows that the transition to
a strongly folded state, presented in ref.\ \cite{LukaCerda}, is a
smooth crossover, not a true transition.  Even when the overall
displacement $\Delta$ is indefinitely smaller than the wrinkling
wavelength $\lambda$, the uniformly wrinkled state is unstable against
a state of localized deformation. 

Furthermore, if the assumption of incompressibility employed
throughout our analysis is relaxed, another non\-zero critical
displacement appears, where a transition from the flat to a wrinkled
state takes place. This buckling occurs at a displacement $\Deltab$
where the in-plane compressional stress $Y (\Deltab/L)$ ($Y$ being the
two-dimensional compression modulus of the sheet), exceeds the
buckling stress of the sheet $2(BK)^{1/2}$. Now, if the system size
$L$ is sufficiently large, we will have $\Deltac<\Deltab$, and the
regime of uniform wrinkling will disappear altogether. The condition
for such a direct flat-to-fold transition \cite{ZhangWitten} is
$L\gtrsim (Y/K)^{1/2} \sim (Y/B)^{1/2}\lambda^2 \sim \lambda^2/t$,
where we have used the scaling of the compression and bending moduli
with the sheet thickness $t$: $Y\sim t$ and $B\sim t^3$. The inclusion
of compressibility renders the transition first-order
\cite{landau:44}, in contrast to the continuous incompressible case
treated by ref.\ \cite{LukaCerda} and in the present work.

The wrinkled state has not been observed in macroscopic Langmuir
monolayers \cite{Lee:2008qe}.  We believe that the theory above
explains this lack of observation. In a typical Langmuir-monolayer
experiment $L\sim 10$ cm and $\lambda\sim 1$ $\mu$m. Hence, a wrinkled
state should get localized already for extremely small displacements,
$\Delta>\Deltac\sim\lambda^2/L\sim 0.1$~\AA. The thickness of the
monolayer is of $1$ nm scale, implying the lack of extended wrinkling
for $L\gtrsim 1$ cm. Thus, we expect that in such experiments there is
a direct transition from flat to folded configurations. Because of the
relevance of compressibility in this experiment, we expect the folds
to form via discontinuous, abrupt events, as observed
\cite{Gopal:2006eu}.  The wrinkled state {\em has} been reported
experimentally \cite{LukaCerda,Leahy:2010zr} in systems that are not
extremely large compared to $\lambda$. As discussed in the Appendix,
the wrinkled state can be stable in systems where
$L\lesssim\lambda^2/\Delta$. Other experiments exhibiting
wrinkling \cite{Huang:2007vn} have strongly constrained lateral
boundaries that likely form a barrier against folding.

The wrinkle-to-fold transition has been applied to describe not only
fluid substrates but also elastic substrates of stiffness $K$.  In
the lowest-order approximation, these two situations are indeed
equivalent.  However, in the next-leading order treated here, the
elastic system becomes different from the fluid system because an
elastic substrate then exerts shear stresses on the sheet, while the
fluid substrate does not.  It may well be that the wrinkled state is
more stable on an elastic substrate than on the liquid substrate
treated above.

It is interesting to compare the present instability with that seen in
an ordinary first-order transition near the spinodal line
\cite{Stanley:1987ve}.  In this case as in ours, the leading-order
contribution to the energy is stabilizing --- favoring zero amplitude
of the new phase.  In both cases the next leading order is
destabilizing.  In both cases one may choose an intensive variable
(\ie pressure) as the control parameter, or an extensive one (\ie
volume or our displacement $\Delta$).  In the phase transition case,
the current phase is metastable.  A nonzero threshold volume of
material must nucleate the new phase in order for the transition to
occur.  In the folding transition, there is likewise a critical
displacement required for any pressure $P$ below the threshold.  As
the control parameter moves to the spinodal line, the new phase
dictated by the lowest-order energy begins to grow spontaneously. The
counterpart of this behavior in our system would be for a wrinkled
phase to grow in amplitude.  However, the above analysis indicates a
different behavior for the case of folding. Rather than growing in
amplitude as dictated by the lowest-order energy, the wrinkled state
collapses into a fold.

The enhanced localization arises because of the
unexpected importance of non-leading energetic terms at indefinitely
small $\Delta$.  An extended state, however small its amplitude, can
be rearranged to produce a localized state of nonvanishing amplitude,
for which nonleading contributions to the energy are significant. (As
we noted, any uniform wrinkle must have vanishing amplitude and hence
vanishing anharmonic energy.)

The localized buckling studied here resembles the buckling of a thick
elastic film that is put under compressive
strain \cite{Kim:2010lq,HohlfeldFold2010}.  In this case as in ours,
linear response suggests an undulating surface. However, higher-order
parts of the energy lead to a localized crease of nonzero slope but
arbitrarily small depth.  An analogous localized instability occurs in
an elastic solid under tensile stress.  Typical models of elastic
solids are linearly unstable to the formation of cavities of
arbitrarily small size \cite{Sivaloganathan:2008rr}.  Our case differs
from these two in that our localized state occupies a region of
diverging size as the threshold is approached.  This ``progressive
localization" bears closer resemblance to that seen in the nonlinear
Schr\"odinger equation for wave-function $\psi$:
\[
- \frac{\pd^2 \psi}{\pd x^2} - |\psi|^2 \psi = E \psi.
\]
Here any negative energy $E$ produces a localized bound state with a
spatial extent that diverges as $E\rightarrow 0$ (albeit 
with no oscillation) \cite{PethickBEC2008}.

\section*{Conclusion}
%====================
\label{sec_conclusion}

The incipient folding instability described above opens several
promising directions.  The energy and in\-cip\-ient shape found here
form the basis for predicting the time dependence of folding.  The
observed abrupt appearance and arrest of folds in time has long been a
puzzle \cite{Gopal:2006eu}. The unexpected simplicity of the energy
versus displacement suggests an underlying symmetry of the system as
yet undiscovered.  Our work on these problems is in progress.

\begin{acknowledgement}
  The authors warmly acknowledge insightful discussions with Ka Yee
  Lee, Luka Pocivavsek, Jin Wang, Enrique Cerda, and Benny
  Davidovitch.  This work was completed at the Aspen Center for
  Physics.  It was supported in part by the US--Israel Binational
  Science Foundation under Grant Number 2006076, and in part by the
  National Science Foundation's MRSEC Program under Award Number DMR
  0820054.
\end{acknowledgement}

\section*{Appendix A. Finite-size effect}
%========================================
\setcounter{equation}{0}
\renewcommand{\theequation}{A.\arabic{equation}}

The analyses in sects.\ \ref{sec_wrinkle} and \ref{sec_fold} have been
restricted to the limit of a very large system size,
$1\ll\kappa^{-1}\ll L$. However, above and sufficiently close to the
wrinkle-to-fold transition (\ie for $\Delta\gtrsim\Deltac=48/L$),
$\kappa$ is arbitrarily small, and the assumption $\kappa L\gg 1$ must
break down. We now treat this limit of very small displacement, where
$(\kappa^{-1},L)\gg 1$ but $\kappa L$ may be finite.

For such minute deformations an `evanescent wave' {\it Ansatz},
\begin{equation}
  h(s) = A\sin (ks) e^{-\kappa|s|},
\label{ansatz1}
\end{equation}
gives the correct qualitative behavior as regards the scaling laws
while simplifying the required integrations. An odd function has been
chosen in eq.\ (\ref{ansatz1}) to avoid a divergent curvature at
$s=0$. We set $k=\kc=1$ and $L=2\pi n$, where $n$ is a large integer
(\ie the system size is compatible with a large integer number of
wrinkling wavelengths). We repeat the same procedure from sect.\ 
\ref{sec_fold} --- substitute eq.\ (\ref{ansatz1}) in eq.\ (\ref{E2}),
calculate $E$ and $\Delta$ to second order in $A^2$, express $A^2$ as
a function of $\Delta$, and substitute it back in the energy to get
$E(\Delta)$. There are two differences in the current calculation: (a)
the integrations in eq.\ (\ref{E2}) are performed over the finite
interval $(-L/2,L/2)$ rather than an infinite one; (b) we expand
the expressions in small $\kappa$ but keep terms that include $\kappa
L$ intact. The resulting energy is
\begin{equation}
  E = 2(1+2\kappa^2)\Delta - \half\kappa \coth\left(\frac{\kappa L}{2}
   \right) \Delta^2.
\label{Efinite}
\end{equation}

Minimization of $E$ with respect to $\kappa$ yields the
following equation:
\begin{equation}
  \kappa L = \frac{\Delta L}{16} f(\kappa L),\ \ 
  f(x) = \frac{\sinh x-x}{\cosh x-1}.
\end{equation}
For $\Delta<\Deltac=48/L$ this equation has only the trivial solution,
$\kappa=0$, corresponding to the wrinkled state. For $\Delta>\Deltac$
a solution of finite $\kappa$ appears, increasing continuously from
zero and corresponding to the folded state.  Thus, sufficiently close
to the transition we may expand eq.\ (\ref{Efinite}) in small $\kappa$
to get
\begin{equation}
  E \simeq E_0 + 4\Delta(1-\Delta/\Deltac)\kappa^2
  +(L^3\Delta^2/720) \kappa^4,
\label{Landau}
\end{equation}
where $E_0=2\Delta-\Delta^2/L$.  This Landau-like energy makes evident
the second-order nature of the transition, $\kappa$ serving as the
order parameter. From eq.\ (\ref{Landau}) we obtain
\begin{eqnarray}
  \kappa &\sim& \left\{
    \begin{array}{ll}
     0,\ \ & \Delta<\Deltac \\
     %\sqrt{30}
     L^{-1}\Delta^{-1/2} (\Delta-\Deltac)^{1/2},\ \ & 
     \Delta\gtrsim\Deltac
    \end{array} \right.
 \nonumber\\
  E -E_0 &\sim& \left\{
    \begin{array}{ll}
     0,\ \ & \Delta<\Deltac \\
     - %(5/4)
     L^{-1}(\Delta-\Deltac)^2,\ \ \ \ \ \ \ \ \ \ & 
     \Delta\gtrsim\Deltac.
    \end{array} \right.
\end{eqnarray}
At $\Delta=\Deltac$ there is a discontinuous jump in $\rmd^2E/\rmd\Delta^2=\rmd P/\rmd\Delta$.
% from $dP/d\Delta=-2L^{-1}=-\Deltac/24$ to
% $dP/d\Delta=-(9/2)L^{-1}=$.

%------------------------------------------------

%--------------------------------------------------

\begin{thebibliography}{99}
  
\bibitem{Bowden1998}
N.\ Bowden, S.\ Brittain, A.\ G.\ Evans, J.\ W.\ Hutchinson,
G.\ W.\ Whitesides, 
Nature {\bf 393}, 146 %--149
(1998). %XX

\bibitem{Sharon:2002kx}
E.\ Sharon, B.\ Roman, M.\ Marder, G.\ Shin, H.\ Swinney,
%Mechanics: Buckling cascades in free sheets - wavy leaves may not
%depend only on their genes to make their edges crinkle.
Nature {\bf 419}, %(6907)
579 %--579 
(2002).

\bibitem{CerdaMaha}
E.\ Cerda, L.\ Mahadevan,
Phys.\ Rev.\ Lett.\ {\bf 90}, 074302 (2003).

\bibitem{Cerda:2004dp}
E.\ Cerda, L.\ Mahadevan, J.\ Pasini,
%The elements of draping.
{\it Proc.\ Natl.\ Acad.\ Sci.\ USA}, {\bf 101}, %(7)
1806 %--1810 
(2004).

\bibitem{Gopal:2006eu}
A.\ Gopal, V.\ Belyi, H.\ Diamant, T.\ A.\ Witten, K.\ Y.\ C.\ Lee,
%Microscopic folds and macroscopic jerks in compressed lipid monolayers.
J.\ Phys.\ Chem.\ B {\bf 110}, %(21)
10220 %--10223
(2006).

\bibitem{Witten:2007ai}
T.\ A.\ Witten,
%Stress focusing in elastic sheets.
Rev.\ Mod.\ Phys.\ {\bf 79}, %(2)
643 %--675
(2007).

\bibitem{Huang:2007vn}
J.\ Huang, M.\ Juszkiewicz, W.\ H.\ de~Jeu, E.\ Cerda, T.\ Emrick, N.\ Menon,
T.\ P.\ Russell,
%Capillary wrinkling of floating thin polymer films.
Science {\bf 317}, %(5838)
650 %--653,
(2007).

\bibitem{Audoly:2008ys}
B.\ Audoly, A.\ Boudaoud,
%Buckling of a stiff film bound to a compliant substrate - part iii:
%Herringbone solutions at large buckling parameter.
J.\ Mech.\ Phys.\ Solids {\bf 56}, %(7)
2444 %--2458 
(2008).

\bibitem{Lee:2008qe}
K.\ Y.\ C.\ Lee,
%Collapse mechanisms of langmuir monolayers.
Ann.\ Rev.\ Phys.\ Chem.\ {\bf 59}, 771 %--791
(2008).

\bibitem{LukaCerda} 
L.\ Pocivavsek, R.\ Dellsy, A.\ Kern, S.\ Johnson, B.\ Lin, K.\ Y.\ C.\ Lee, 
E.\ Cerda, 
Science {\bf 320}, 912 (2008).

\bibitem{Py:2009uq}
C.\ Py, P.\ Reverdy, L.\ Doppler, J.\ Bico, B.\ Roman, C.\ N.\ Baroud,
%Capillarity induced folding of elastic sheets.
Eur.\ Phys.\ J.\ Special Topics {\bf 166}, 67 %--71
(2009).

\bibitem{BennyPRE}
B.\ Davidovitch,
Phys.\ Rev.\ E {\bf 80}, 025202 (2009).

\bibitem{Huang:2010sp}
J.\ Huang, B.\ Davidovitch, C.\ D.\ Santangelo, T.\ P.\ Russell, N.\ Menon,
%Smooth cascade of wrinkles at the edge of a floating elastic film.
Phys.\ Rev.\ Lett.\ {\bf 105}, 038302 (2010).

\bibitem{Holmes:2010fk}
D.\ P.\ Holmes, A.\ J.\ Crosby,
%Draping Films: A Wrinkle to Fold Transition.
Phys.\ Rev.\ Lett.\ {\bf 105}, 038303 (2010).

\bibitem{Leahy:2010zr}
B.\ D.\ Leahy, L.\ Pocivavsek, M.\ Meron, K.\ L.\ Lam, D.\ Salas, P.\ J.\ Viccaro,
K.\ Y.\ C.\ Lee, B.\ Lin,
%Geometric stability and elastic response of a supported nanoparticle film.
Phys.\ Rev.\ Lett. {\bf 105}, 058301 (2010).

\bibitem{landau:44}
L.\ D.\ Landau, L.\ M.\ Lifshitz, 
{\it Theory of Elasticity} (Pergamon, New York, 1986).

\bibitem{Milner}
S.\ T.\ Milner, J.-F.\ Joanny, and P.\ Pincus,
Europhys.\ Lett.\ {\bf 9}, 495 (1989).

\bibitem{ZhangWitten}
Q.\ Zhang and T.\ A.\ Witten,
Phys.\ Rev.\ E {\bf 76}, 041608 (2007).

\bibitem{Stanley:1987ve}
H.\ E.\ Stanley,
{\it Introduction to Phase Transitions and Critical Phenomena}
(Oxford University Press, New York, 1987).

\bibitem{Kim:2010lq}
J.\ Kim, J.\ Yoon, R.\ C.\ Hayward,
%Dynamic display of biomolecular patterns through an elastic creasing
%instability of stimuli-responsive hydrogels.
Nat.\ Mater.\ {\bf 9}, %(2)
159 %--164
(2010).

\bibitem{HohlfeldFold2010}
E.\ Hohlfeld, L.\ Mahadevan,
%Unfolding the sulcus.
e-print arXiv:1008.0694, 2010.

\bibitem{Sivaloganathan:2008rr}
J.\ Sivaloganathan, S.\ J.\ Spector,
%Energy minimising properties of the radial cavitation solution in
%incompressible nonlinear elasticity
J.\ Elasticity {\bf 93}, %(2)
177 %--187}
(2008).

\bibitem{PethickBEC2008}
C.\ Pethick,
{\it Bose-Einstein Condensation in Dilute Gases}
(Cambridge University Press, New York, 2008).


\end{thebibliography}
\end{document}